\documentclass{stacs}

\usepackage{algorithmic}
\usepackage{algorithm}

\usepackage{graphicx}

\usepackage{amsmath}
\usepackage{amsthm}
\usepackage{amsbsy}
\usepackage{amssymb}
\usepackage{amsfonts}
\usepackage{setspace}
\usepackage[dvips]{lscape}

\DeclareMathOperator*{\argmin}{argmin}
\DeclareMathOperator*{\rank}{rank}

\DeclareMathOperator*{\card}{card}

\DeclareMathOperator*{\dist}{dist}

\DeclareMathOperator*{\rad}{rad}

\DeclareMathOperator*{\diam}{diam}

\DeclareMathOperator*{\vertex}{vert}
\DeclareMathOperator*{\cost}{cost}

\DeclareMathOperator*{\spanSpace}{span}
\DeclareMathOperator*{\proj}{proj}

\DeclareMathOperator*{\vol}{vol}
\DeclareMathOperator*{\volG}{vol'}
\DeclareMathOperator*{\zOptV}{z_v}
\DeclareMathOperator*{\zOptD}{z_d}
\DeclareMathOperator*{\zOptR}{z_r}

\begin{document}

% \title[short title]{title}
\title{Quantifying Homology Classes II: Localization and Stability}

% \author[ref]{Short author}{Author}
\author[lab1]{}{Chao Chen}
\author[lab1]{}{Daniel Freedman}
\address[lab1]{Rensselaer Polytechnic Institute
\newline 110 8th street, Troy, NY 12180, U.S.A.}  %required
\email[C. Chen]{chenc3@cs.rpi.edu} %optional
\email[D. Freedman]{freedman@cs.rpi.edu}  %optional

%% mandatory lists of keywords and classifications:
% \titlecomment{OPTIONAL comment concerning the title, \eg, if a variant
% or an extended abstract of the paper has appeared elsewehere}
%%%%%%%%%%%%%%%%%%%%%%%%%%%%%%%%%%%%%%%%%%%%%%%%%%%%%%%%%%%%%%%%%%%%%%%%%%%

%% the abstract has to PRECEDE the command \maketitle:
%% be sure not to issue the \maketitle command twice!

\begin{spacing}{0.95}

\begin{abstract}
  \noindent In the companion paper \cite{ChenF1}, we measured homology classes and computed the optimal homology basis. This paper addresses two related problems, namely, localization and stability. We localize a class with the cycle minimizing a certain objective function. We explore three different objective functions, namely, volume, diameter and radius. We show that it is NP-hard to compute the smallest cycle using the former two. We also prove that the measurement defined in \cite{ChenF1} is stable with regard to small changes of the geometry of the concerned space.
\end{abstract}

\maketitle
\vspace{-0.2 in}
%%%%%%%%%%%%%%%%%%%%%%%%%%%%%%%%%%%%%%%%%%%%%%%%%%%
\section{Introduction}
The problem of computing the topological features of a space has recently drawn much attention from researchers in various
fields, such as high-dimensional data analysis  \cite{Carlsson05,Ghrist}, graphics \cite{EricksonH04,CarnerJGQ05}, networks \cite{SilvaG06} and computational biology \cite{AgarwalEHW06,Cohen-SteinerEM06}. Topological features are often preferable to purely geometric features, as they are more qualitative and global, and tend to be more robust. If the goal is to characterize a space, therefore, features which incorporate topology seem to be good candidates. 
In this paper, the topological features we use are homology groups over $\mathbb{Z}_2$, due to their ease of computation. (Thus, throughout this paper, all the additions are mod 2 additions.)

In the companion paper \cite{ChenF1}, we addressed the problems of measuring homology classes and finding a concise representation (a homology basis) of the homology group. We defined a {\it size} of a homology class as the radius of the smallest geodesic ball carrying the class, using terminology from relative homology. %formally, 
%\begin{equation*}
%S(h)=\min_{ B_p^r } r \quad s.t.\quad \phi_{B_p^r}^*(h)=\mathsf{B}_d( K , B_p^r ).
%\end{equation*}
We also computed the {\it optimal homology basis} which is the homology basis whose elements' size have the minimal sum.
%, formally,
%\begin{equation*}
%\mathcal{H}_d=\argmin_{\{h_1,...,h_{\beta_d}\}}\sum_{i=1}^{\beta_d} S(h_i),s.t.
%\dim(\{h_1,...,h_{\beta_d}\})=\beta_d.
%\end{equation*}

This paper addresses two problems that are natural byproducts of the ideas in \cite{ChenF1}.
\paragraph{\bf Problem 1: Localization.} 
In the companion paper \cite{ChenF1}, we localized each class in the optimal homology basis with its {\it localized-cycle}, which is the representative cycle carried by the smallest geodesic ball carrying the class. That was necessary for the algorithm to work, but was otherwise not an object of interest. However, the use of localized-cycle suggests an interesting problem: given a natural criterion of the size of a cycle, can we find the smallest cycle in a class? We will see that the companion paper implicitly provided one such criterion, but there are other criteria as well. 

\begin{figure}[hbtp]
    \vspace{-0in}
    \centerline{
    \begin{tabular}{c|c}
		\includegraphics[width=0.25\textwidth]{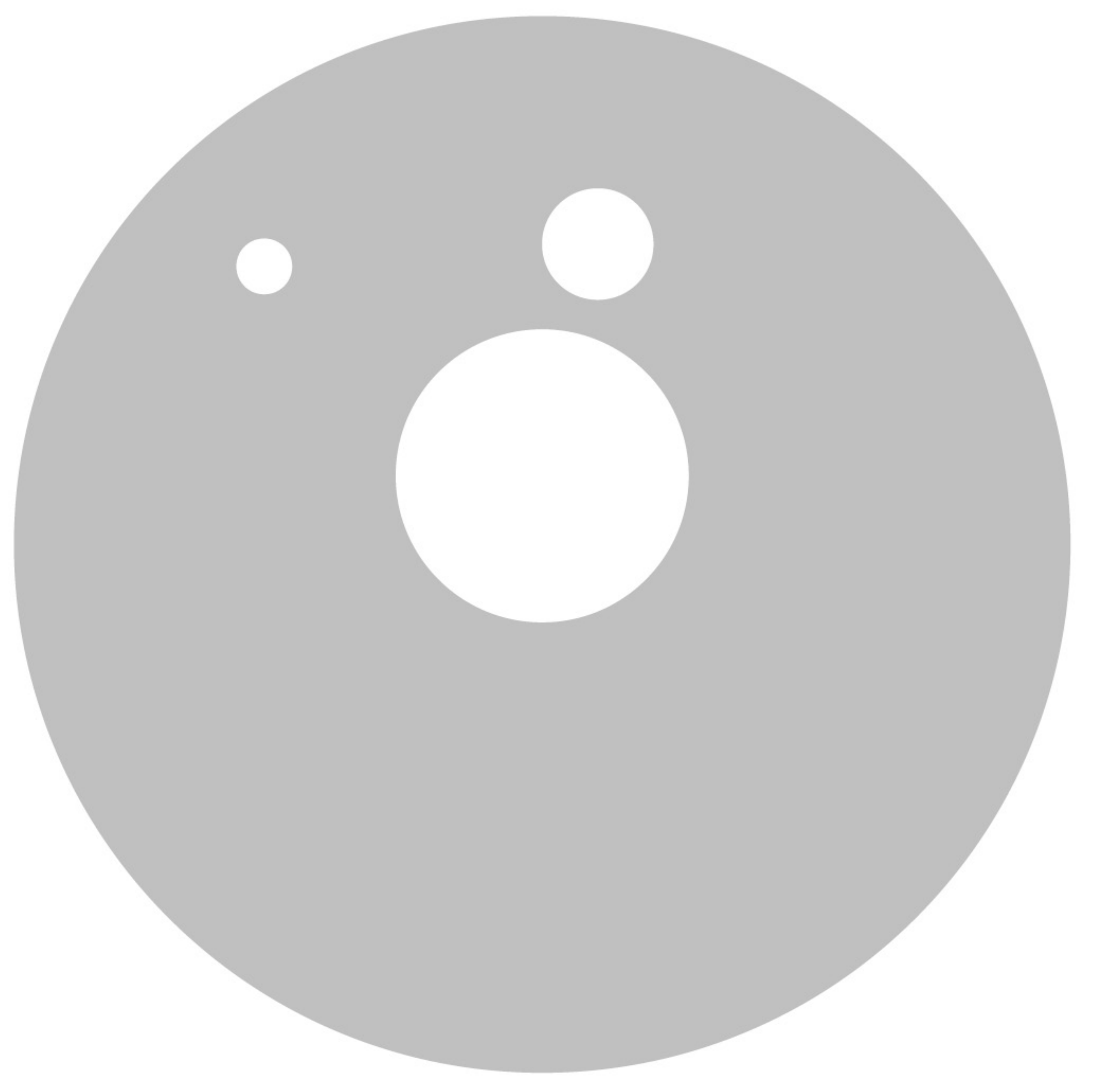} &
		\includegraphics[width=0.25\textwidth]{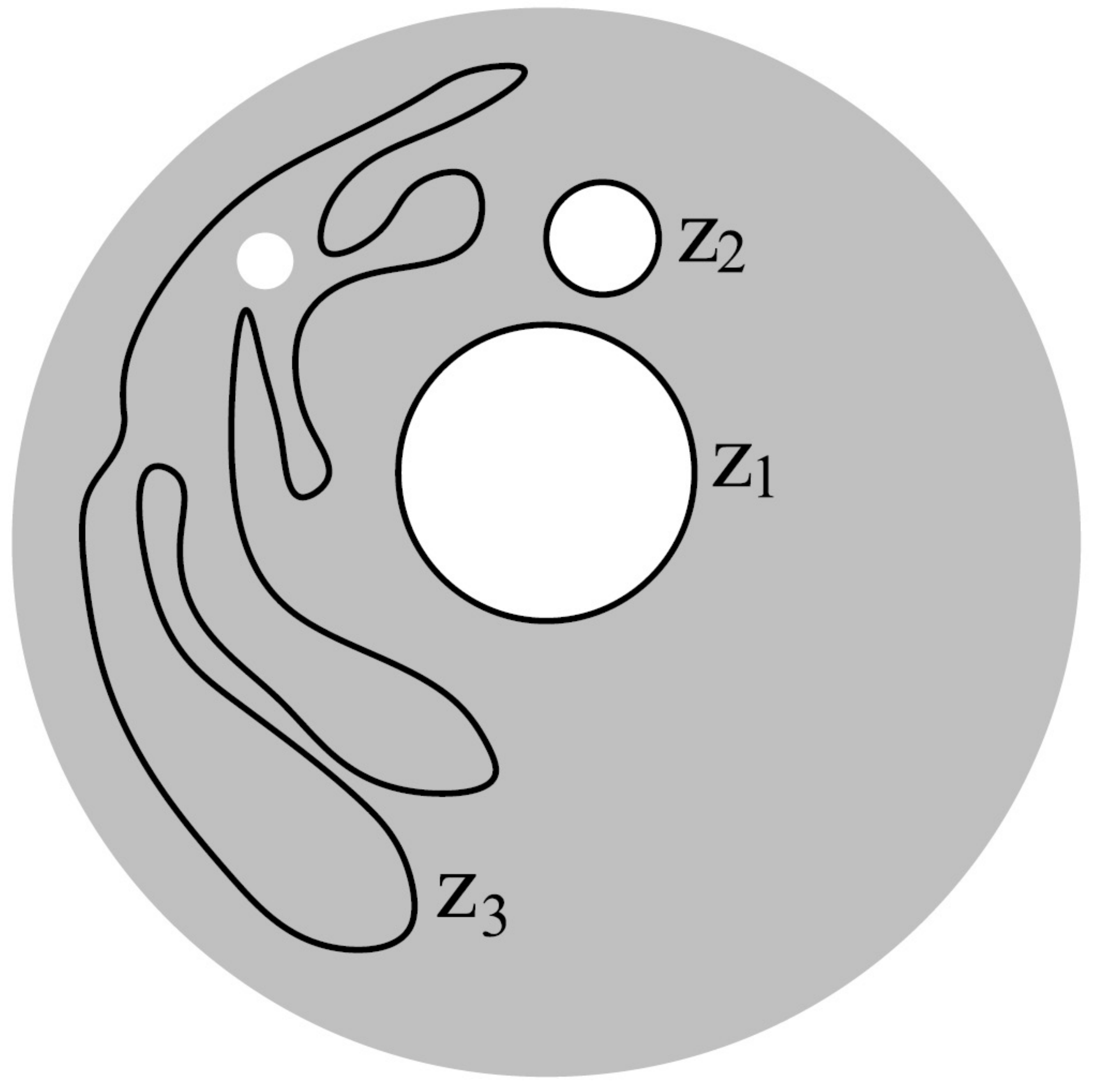}
    \end{tabular}}
    \vspace{-0in}
    \caption{\small\bf Left: the concerned space is a disk with three holes.
    \newline Right: cycles $z_1$ and $z_2$ are well-localized; $z_3$ is not. 
		}
    \label{fig:motivation}
    \vspace{-0in}
\end{figure}

The criterion should be deliberately chosen so that the corresponding smallest cycle is concise in not only mathematics but also intuition. Such a cycle is a ``well-localized'' representative cycle of its class. For example, in Figure \ref{fig:motivation}, the cycles $z_1$ and $z_2$ are well-localized representatives of their respective homology classes; whereas $z_3$ is not.
This problem of localizing classes with well-localized cycles has potential applications in various field such as high-dimensional data analysis  \cite{Carlsson05,Ghrist}, graphics \cite{WoodHDS04,EricksonH04,CarnerJGQ05}, CAD \cite{DeyLS07}, shape study \cite{CarlssonZCG05}, etc. A review of approaches of this problem will be given later in this section. 

In this paper, we will look at three such criteria, i.e.~the volume, diameter and radius of a cycle. We will show that computing the smallest cycle using the former two are NP-hard, whereas it is polynomial to compute using the latter, which corresponds naturally to the homology class size measure used in the companion paper.

\paragraph{\bf Problem 2: Stability.} We also address another open question suggested by the companion paper. The size measure of homology classes and the optimal homology basis defined in that paper depend on the (discrete) metric on the simplicial complex under consideration. A natural question is: is the measure and the basis stable with respect to small changes in metric? In Section \ref{sec:stability} of this paper, we show that under a suitable definition of stability, the answer is yes. This provides theoretical justification for our work in the companion paper \cite{ChenF1}.

\paragraph{\bf Restrictions.} Furthermore, as in the companion paper, we make two additional requirements on the solution of the aforementioned problem. First, the solution ought to be computable for topological spaces of arbitrary dimension. Second the solution should not require that the topological space be embedded, for example in a Euclidean space; and if the space is embedded, the solution should not make use of the embedding.  These requirements are natural from the theoretical point of view, but may also be justified based on the following applications:
\begin{itemize}
\item In machine learning, it is often assumed that the data lives in a manifold whose dimension is much smaller than the dimension of the embedding space.

\item In the study of shape, it is common to enrich the shape with other quantities, such as curvature, or color and other physical quantities. This leads to high dimensional manifolds (e.g, 5-7 dimensions) embedded in high dimensional ambient spaces \cite{CarlssonZCG05}. 
\end{itemize}

\paragraph{\bf Related Works.}
Using Dijkstra's shortest path algorithm, Erickson and Whittlesey \cite{EricksonW05} localized a 1-dimensional homology class with its shortest cycle. They defined the {\it shortest homology basis} as the set of $\beta$ linearly independent homology classes, such that lengths of their shortest representative cycles have the minimal sum. They provided a greedy algorithm to localize classes in this basis with their shortest cycles.\footnote{Note that their polynomial algorithm can only localize classes in the shortest homology basis. In fact, we will show in Section \ref{sec:vol} that it is NP-hard to localize an arbitrary given class with the shortest cycle.} The authors also showed how the idea carries over to finding the optimal generators of the first fundamental group, though the proof is considerably harder in this case. For completeness, we refer to some related works which compute a single or a set of non-trivial cycles satisfying certain topological and geometrical restrictions on 2-manifolds \cite{EricksonH04,Kutz06,ChambersVELW06,VerdiereE06}.

Some researchers concentrate on 1-dimensional cycles closely related to handles which are much more meaningful in low dimensional applications such as graphics and CAD. Given a 2-manifold embedded in $S^3$, Dey et al.~\cite{DeyLS07}  computed these handle-related cycles by computing the deformation retractions of the two components of the the embedding space bounded by the given 2-manifold. Their work facilitates handle detection in real applications. However, the computed 1-cycles are not guaranteed to be geometrically concise. Guskov and Wood \cite{GuskovW01,WoodHDS04} detected small handles of a 2-manifold using the Reeb graph of the manifold.

All of the aforementioned works are restricted to low-dimensional manifolds. Zomorodian and Carlsson \cite{ZomorodianC07} took a different approach to solving the localization problem for general dimension. Their method starts with a topological space and a cover, which is a set of spaces whose union contains the original space. They computed a homology basis and localized classes of it, using tools from algebraic topology and persistent homology. However, both the quality of the localization and the complexity of the algorithm depend strongly on the choice of cover; there is, as yet, no suggestion of a canonical cover.

\paragraph{\bf Contributions.}
We explore the localization problem using different size definitions of a cycle, namely, the {\it volume}, {\it diameter} and {\it radius}.
\begin{itemize}
\item We prove that it is NP-hard to localize a class with its smallest cycle in terms of the volume and the diameter, by reduction from MAX-2SAT-B and MCCP, respectively. 

\item The cycle with the minimal radius is literally the localized-cycle as defined in the companion paper. We show that although it may not be perfectly well-localized, it is a 2-approximation of the cycle with the minimal diameter. A polynomial algorithm to compute this cycle is provided.\footnote{In the companion paper, a polynomial algorithm is provided to compute localized-cycles for classes in the optimal homology class. In this paper, we extend it to localizing any given class.}
\end{itemize}

For the stability problem, we prove that (1) the size of homology classes and (2) the subgroup filtration computed from the optimal homology basis are both stable with regard to small changes of the geometry of the concerned space.

%%%%%%%%%%%%%%%%%%%%%%%%%%%%%%%%%%%%%%%%%%%%%%%%%%%
%%%%%%%%%%%%%%%%%%%%%%%%%%%%%%%%%%%%%%%%%%%%%%%%%%%

\section{Localization}
\label{sec:localization}
In this section, we address the localization problem. For ease of computation, we restrict our work to homology groups over $\mathbb{Z}_2$ field. Throughout this paper, all the additions are mod 2 additions. When we talk about a $d$-dimensional chain, $c$, we refer to either a collection of $d$-simplices, or a $n_d$-dimensional vector over $\mathbb{Z}_2$ field, whose non-zero entries corresponds to the included $d$-simplices. Here $n_d$ is the total number of $d$-simplices in the given complex. The relevant background in homology can be found in \cite{Munkres84}.

We formalize the localization problem as a combinatorial optimization problem: Given a simplcial complex $K$, compute the representative cycle of a given homology class minimizing a certain objective function. Formally, given an objective function defined on all the cycles, $\cost:\mathsf{Z}_d(K)\rightarrow \mathbb{R}$, we want to localize a given class with its {\it optimally localized cycle},
\begin{eqnarray*}
z_{opt}(h)=\argmin_{z\in h}\cost(z).
\end{eqnarray*}
In general, we assume the class $h$ is given by one of its representative cycles, $z_0$.

In this paper, we explore three options of the objective function $\cost(z)$, i.e.~the {\it volume}, {\it diameter} and {\it radius}, in the following three subsections.
\subsection{Volume}
\label{sec:vol}
The first choice of the objective function is the volume.
\begin{definition}[Volume]
\label{def:vol}
The volume of a cycle is the number of its simplices, $\vol(z)=\card(z)$.
\end{definition}
For example, the volume of a 1-dimensional cycle, a 2-dimensional cycle and a 3-dimensional cycle are the numbers of their edges, triangles and tetrahedra, respectively. A cycle with the smallest volume, denoted as $z_v$, is consistent to a ``well-localized'' cycle in intuition. Its 1-dimensional version, the shortest cycle of a class, has been studied by researchers \cite{EricksonW05,WoodHDS04,DeyLS07}. However, we prove in Theorem \ref{thm:optVol} that computing $\zOptV$ of $h$ is NP-hard.\footnote{Erickson and Whittlesey \cite{EricksonW05} localized 1-dimensional classes with their shortest representative cycles. Their polynomial algorithm can only localize classes in the shortest homology basis, not arbitrary given classes.}

More generally, we can extend the the volume to be the sum of the weights assigned to simplices of the cycle, given an arbitrary weight function, $w:K\rightarrow \mathbb{R}$, defined on all the simplices of $K$, formally,
\begin{equation*}
\volG(z)=\sum_{\sigma\in z}w(\sigma).
\end{equation*}
%where $w:K\rightarrow \mathbb{R}$ assigns a real value to each simplex in the simplicial complex. 
Theorem \ref{thm:optVol} implies that computing $\zOptV$ using this general volume definition is still NP-hard, because Definition \ref{def:vol} is in fact a special case of this general definition (when $w(\sigma)=1$, $\forall \sigma\in K$). 

We prove by theorem \ref{thm:optVol} by reduction from the NP-hard problem MAX-2SAT-B \cite{PapadimitriouY88}. 
\begin{problem}[MAX-2SAT-B]
Given $N$ literals, $x_1$ to $x_N$, and $M$ different clauses, $c_1$ to $c_M$, each in the form of $(x_i \vee x_j)$, $(\neg x_i \vee x_j)$, $(x_i \vee \neg x_j)$ or $(\neg x_i \vee \neg x_j)$. Each literal appears in at most $B$ clauses. Find an assignment assigning boolean values to all the literals such that the number of satisfied clauses are maximized.
\end{problem}
\begin{theorem}
\label{thm:optVol}
Computing $\zOptV$ for a given $h$ is NP-hard.
\end{theorem}
\proof
Given an input of MAX-2SAT-B, we construct in polynomial time an input of the problem of computing $\zOptV$ for a given $h$, denoted as the Opt-Vol problem for short. We will later show that the answer of Opt-Vol gives us the answer of MAX-2SAT-B. The input simplcial complex for Opt-Vol consists of $N$ literal tubes and $M$ 1-dimensional clause cycles.
\begin{enumerate}
\item {\bf Literal Tubes.}
For each literal $x_i$, we construct a literal tube $T_i$. Denote $h_i$ as the 1-dimensional class carried by $T_i$. $T_i$ is made fat and long yet with two short end cycles, such that a representative cycle of $h_i$ with no more than $L$ length can only be one of the two end cycles of the tube. (The parameter $L$ will be given later.) See Figure \ref{fig:FatLongTube} (Left) for an example, technical details of the construction are omitted because of the space limitation. Let the two end cycles of $T_i$ be $z_i$ and $z_i'$, corresponding to $x_i$ and $\neg x_i$, respectively. Let all the $2N$ end cycles of the $N$ literal tubes have equal length.
\begin{figure}[hbtp]
    \centerline{
    \begin{tabular}{c|c}
		\includegraphics[width=0.5\textwidth]{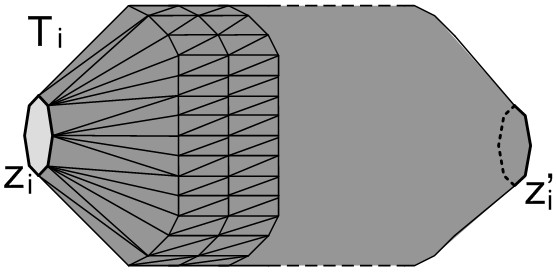} &
		\includegraphics[width=0.4\textwidth]{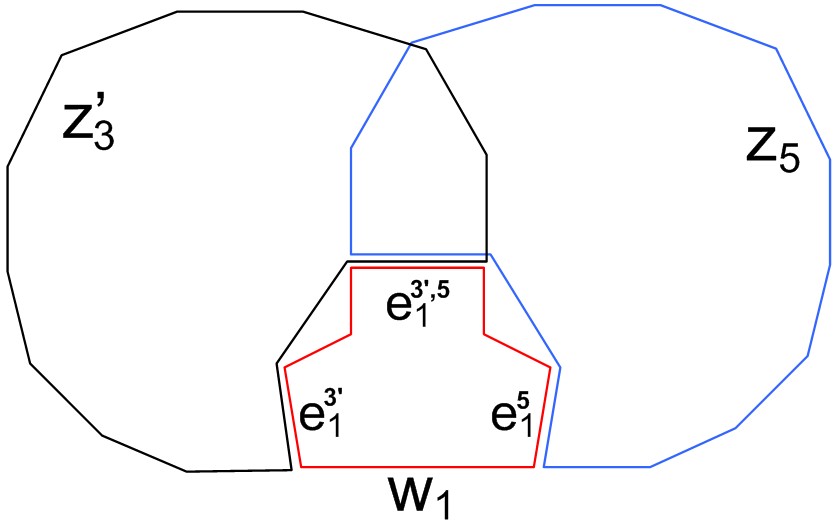}
    \end{tabular}}
    \caption{\small\bf Left: A literal tube $T_i$ is made fat and long yet with two short end cycles, $z_i$ and $z_i'$.
    \newline Right: Example of shared edges. The clause cycle $w_1$ corresponds to the clause $c_1=(\neg x_3 \vee x_5)$. The red, black and blue cycles are $w_1$, $z_3'$ and $z_5$, respectively. Although there is unwanted edge intersections in the figure, the simplicial complex can be studied without embedding, and thus, is well defined.}
    \label{fig:FatLongTube}
\end{figure}
%\begin{figure}[hbtp]
%    \centering
%		\includegraphics[width=0.4\textwidth]{figures/SharingEdges2.jpg}
%    \caption{\small\bf The clause cycle $w_1$ corresponding to the clause $c_1=(\neg x_3 \vee x_5)$. The red, black and blue cycles are $w_1$, $z_3'$ and $z_5$, respectively.}
%    \label{fig:SharingEdges}
%\end{figure}

\item {\bf Clause Cycles and Shared Edges.}
For each clause $c_i$, we construct a 1-dimensional cycle $w_i$. All these clause cycles have equal length. For a clause $c_i = (x_a \vee x_b)$, pick three edges of $w_i$ as shared edges, $e_i^a$, $e_i^b$ and $e_i^{a,b}$. Let cycle $w_i$ and the end cycle $z_a$ share $e_i^a$, $w_i$ and $z_b$ share $e_i^b$. Let $e_i^{a,b}$ be shared by the three cycles, $w_i$, $z_a$ and $z_b$. If $c_i = (\neg x_a \vee x_b)$, substitute the end cycle $z_a'$ for $z_a$  in the aforementioned sharing, and denote the three shared edges as $e_i^{a'}$, $e_i^{b}$ and $e_i^{a',b}$. The case of $(x_a \vee \neg x_b)$ and $(\neg x_a \vee \neg x_b)$ can be addressed similarly. See Figure \ref{fig:FatLongTube} (Right) for an example. 

We deliberately put five extra edges between the three shared edges of $w_i$. Within each end cycle, we put at least one extra edge between any two shared edges. These extra edges make sure that each vertex of the clause cycles and the end cycles belong to at most one shared edge. This way we can avoid discussions about the validation of the constructed simplicial complex, e.g.~two edges sharing two vertices, etc.

\item {\bf Length of Cycles.}
For a clause cycle $w_i$, as shown in Figure \ref{fig:FatLongTube} (Right), we have three shared edges and five extra edges. Its length is eight. For an end cycle $z_i$ (or $z_i'$), it shares two edges with each involved clause. Since $x_i$ (or $\neg x_i$) appears at most $B$ times, $z_i$ (or $z_i'$) has at most $2B$ shared edges. We put at least one extra edge between any two shared edges. Let each end cycle has $4B$ edges.
\end{enumerate}

Next, for the constructed simplicial complex, we ask Opt-Vol to localize the homology class $h=\sum_{i=1}^N h_i + \sum_{i=1}^M [w_i]$ with the cycle with the smallest volume, which is, the shortest cycle. We may need a cycle to represent $h$ as the input. The cycle $z_0=\sum_{i=1}^N z_i + \sum_{i=1}^M w_i$ will work. Next, we show that there is a bijection between the set of boolean assignments of the literals and the set of representative cycles of $h$ with no more than $4NB+8M$ edges. Under this bijection, a boolean assignment satisfies $k$ clauses if and only if its corresponding cycle has $4NB+8M-2k$ edges.

It is not hard to see that localizing $h$ is equivalent to finding a representative cycle for each $h_i$ and sum these cycles up with all the clause cycles, $w_i$. Recall that we make each tube fat and long such that a representative cycle of $h_i$ with no more than $L$ edges can only be one of its end cycles, $z_i$ or $z_i'$. Let the parameter $L=4NB+8M$. Then, any representative cycle of $h$ with no more than $4NB+8M$ edges can only be the sum of the $M$ clause cycles and $N$ end cycles, coming from the $N$ tubes, formally, $\sum_{i=1}^M w_i+\sum_{i=1}^N \tilde{z}_i$, $\tilde{z}_i$ is either $z_i$ or $z_i'$. For a boolean assignment $X$, we construct the bijection such that in its corresponding cycle $z$, $\tilde{z}_i=z_i$ iff $x_i=True$ and $\tilde{z}_i=z_i'$ iff $x_i=False$.

For each boolean assignment, $X$, and the corresponding representative cycle, $z$. A clause $c_i$ is satisfied by $X$ if and only if one of the three shared edges of $w_i$ appear in $z$. $c_i$ is NOT satisfied if and only if all three shared edges of $w_i$ appear. We can see this by enumerating all possibilities:
\begin{enumerate}
\item $c_i$ is satisfied if and only if one of the following three cases happen.
\begin{itemize}
\item $x_a=True$ and $x_b=True$. $w_i$, $z_a$ and $z_b$ are all chosen. In this case, edge $e_i^a$ and $e_i^b$ are canceled in $z$ because of the mod 2 addition. Edge $e_i^{a,b}$ is NOT because we count it three times;
\item $x_a=True$ and $x_b=False$. $w_i$ and $z_a$ are chosen. $z_b$ is NOT. Edge $e_i^a$ and $e_i^{a,b}$ are canceled. Edge $e_i^b$ is NOT canceled.
\item $x_a=False$ and $x_b=True$. Similar to the second case. 
\end{itemize}
\item $c_i$ is NOT satisfied if and only if $x_a=False$ and $x_b=False$, which means $z_a$ and $z_b$ are not chosen. Since $w_i$ is always chosen, all the three shared edges are not canceled.
\end{enumerate}
In $z$, $M$ clause cycles contribute $8M$ edges and $N$ end cycles contribute $4NB$ edges. Consequently,
$k$ of the $M$ clauses are satisfied by $X$ if and only if $2k$ shared edges are canceled, that is, $z$ has $4NB+8M-2k$ edges.

With this bijection, it is not hard to see that a cycle with the minimal number of edges uniquely corresponds to a boolean assignment satisfying the maximal number of clauses.
\qed

\subsection{Diameter}
\label{sec:diam}
When it is NP-hard to compute $\zOptV$, one may resort to the geodesic distance between elements of $z$, denoted as $\dist(p,q)$. As we will deal with a simplicial complex $K$, it is natural to introduce a discrete metric, that is, a length on each edge. Define $\dist(p,q)$ as the length of the shortest path connecting the two vertices $p$ and $q$, in the 1-skeleton of $K$. 

The second choice of the objective function is the diameter.
\begin{definition}[Diameter]
\label{def:diam}
The diameter of a cycle is the diameter of its vertex set, $\diam(z)=\diam(\vertex(z))$, in which the diameter of a set of vertices is the maximal geodesic distance between them, formally, 
\begin{equation*}
\diam(S)=\max_{p,q\in S}\dist(p,q).
\end{equation*}
\end{definition}
Intuitively, a representative cycle of $h$ with the minimal diameter, denoted $z_d$, is the cycle whose vertices are as close to each other as possible. The intuition will be further illustrated in Section \ref{sec:rad} by comparison against other criteria. We prove in Theorem \ref{thm:optDiam} that computing $\zOptD$ of $h$ is NP-hard, by reduction from the NP-hard {\it Multiple-Choice Cover Problem} (MCCP) by Arkin and Hassin \cite{ArkinH00}.
\begin{problem}[Multiple-Choice Cover Problem]
Given $n$ vertices, $V=\{v_1,v_2,...,v_n\}$ in the Euclidean plane, $\mathbb{E}$, let $d(v_i,v_j)$ be the Euclidean distance between any two vertices $v_i$ and $v_j$. Let $S_1,S_2,...,S_m\subseteq V$ be disjoint subsets of $V$ such that $\bigcup_{i=1}^m S_i=V$. Find a {\it cover} $C\subseteq V$ containing one and exactly one vertex from each subset $S_i$, whose diameter, $\diam(C)$, is minimized.\footnote{The original MCCP problem by Arkin and Hassin only requires the cover to have nonempty intersection with each subset $S_i$. We revise the problem to facilitate our proof, without influencing the NP-hardness. The reason is the optimal result of the revised problem is definitely an optimal result of the original problem.}
\end{problem}
\begin{theorem}
\label{thm:optDiam}
Computing $\zOptD$ for a given $h$ is NP-hard.
\end{theorem}
\proof
We present a polynomial algorithm transforming an input of MCCP into an input of the Opt-Diam problem, namely, computing $z_d$ for a given $h$. We will show that the solution of Opt-Diam gives us the solution of MCCP. As part of the input of Opt-Diam, the constructed simplicial complex $K$ consists of $m$ tubes, $T_1$,...,$T_m$, as well as extra edges connecting vertices.  

\begin{figure}[hbtp]
    \centerline{
    \begin{tabular}{l|r}
    \includegraphics[width=0.4\textwidth]{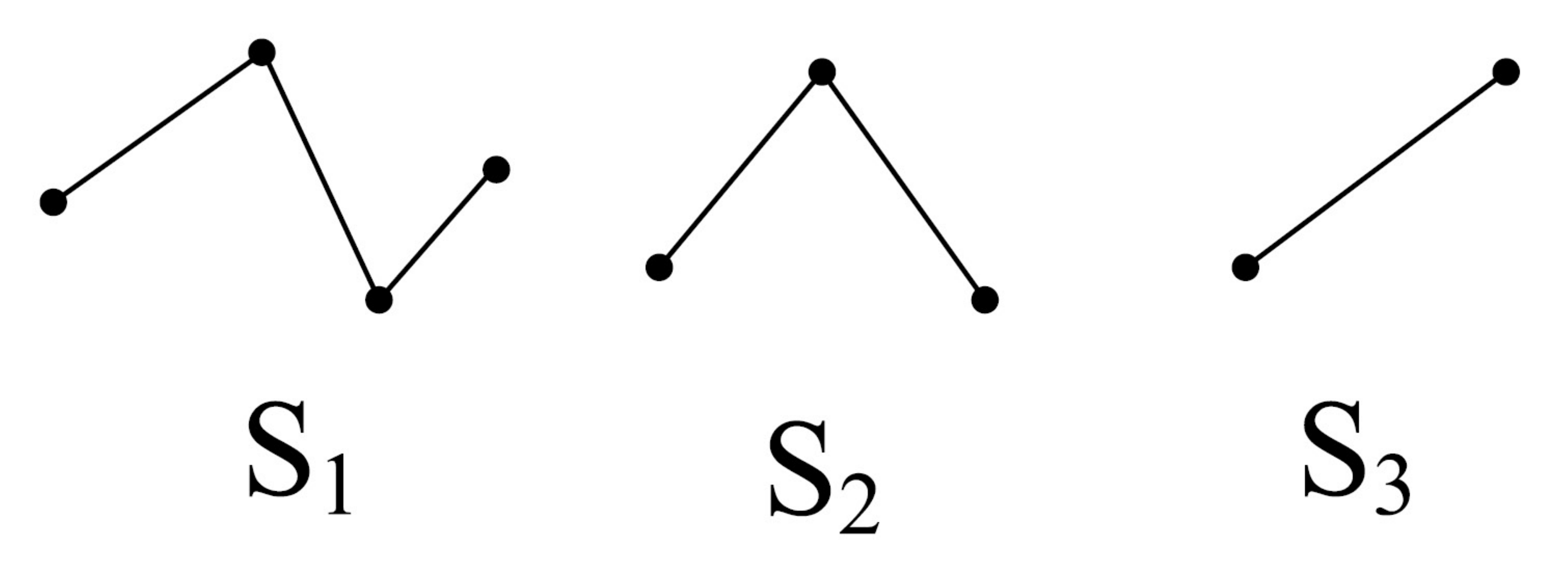} &
		\includegraphics[width=0.25\textwidth]{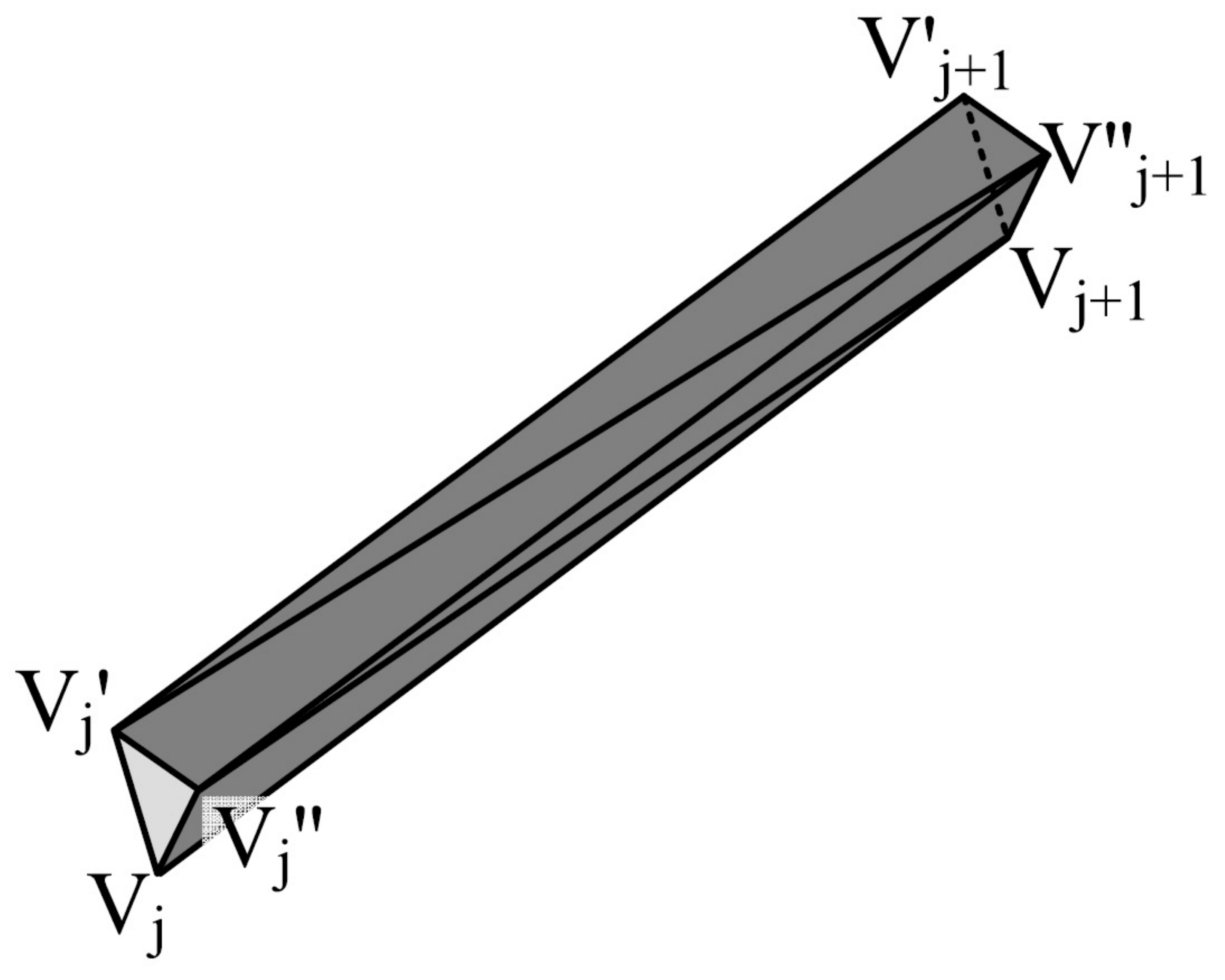}
    \end{tabular}}
    \vspace{-0in}
    \caption{\small\bf Left: An input of MCCP: 3 disjoint vertex subsets in Euclidean plane, $S_1$, $S_2$ and $S_3$. The simple paths, $\xi_1$, $\xi_2$ and $\xi_3$, are also shown, although they are not part of the input.
    \newline Right: A triangular cylinder corresponding to the edge $v_jv_{j+1}$.}
    \label{fig:MCCPPts}
\end{figure}

For each vertex subset $S_i$, we find a simple path in the Euclidean plane, $\mathbb{E}$, going though each vertex of $S_i$ once without self-intersection, $\xi_i=(v_1,v_2,...,v_{\card(S_i)})$, which contains $\card(S_i)-1$ edges. The edge lengths are the same as the Euclidean distances between corresponding vertices in $\mathbb{E}$. See Figure \ref{fig:MCCPPts} (Left).
We construct a a slender threadlike tube $T_i$, which carries the path $\xi_i$. $T_i$ has $3\card(S_i)$ vertices, $S_i\cup S_i' \cup S_i''$, where $S_i'=\{v_1',v_2',...,v_{\card(S_i)}'\}$ and $S_i''=\{v_1'',v_2'',...,v_{\card(S_i)}''\}$. For any $j$, $v_j'$ and $v_j''$ live very close to $v_j$. Corresponding to the $\card(S_i)-1$ edges in $\xi_i$, $T_i$ consists of $\card(S_i)-1$ triangular cylinder concatenated together.\footnote{By a triangular cylinder we mean the surface of a 3-prism with the two end triangles missing. To facilitate the concatenation, corresponding edges of the end triangles may not be parallel to each other, as in a standard 3-prism.} Each edge $v_jv_{j+1}$ corresponds to a triangular cylinder with vertices $v_j$, $v_j'$, $v_j''$, $v_{j+1}$, $v_{j+1}'$ and $v_{j+1}''$. In the triangular cylinder, the short edges are very short, say, no longer than $\epsilon$. The long edges have the length similar to the length of edge $v_jv_{j+1}$. See Figure \ref{fig:MCCPPts} (Right) for one such triangular cylinder.

\begin{figure}[hbtp]
    \centering
    \includegraphics[width=0.65\textwidth]{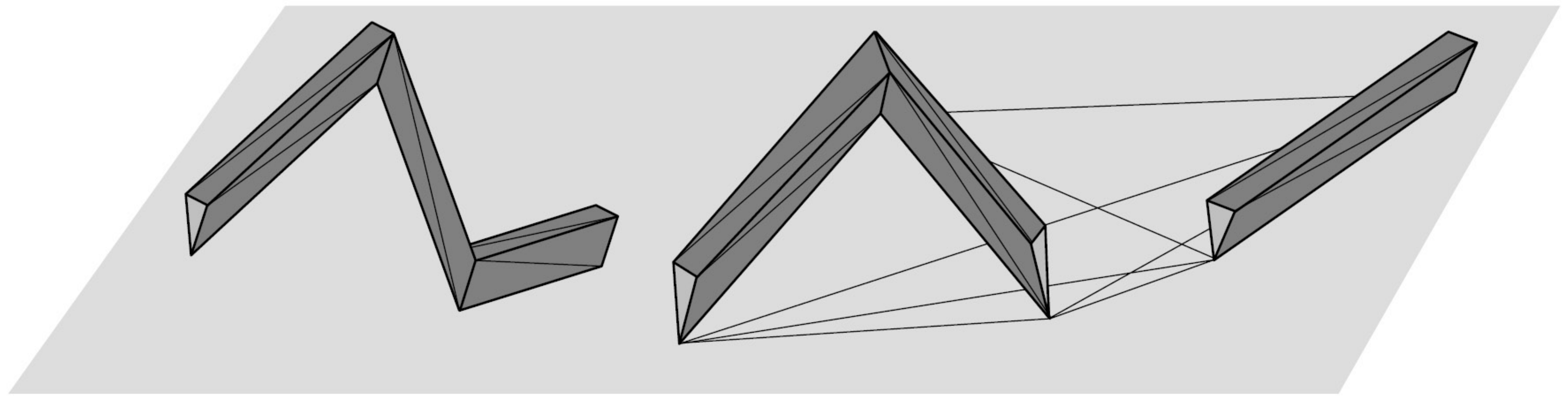}
    \caption[Diam-Opt is NP-Hard: The Input for Diam-Opt.]{\small\bf The constructed simplicial complex $K$. For simplicity, we only include new edges connecting vertices in $S_2\cup S_3$. The rest edges are in $K$ but omitted in this figure.}
    \label{fig:MCCPSpace}
\end{figure}

We construct the simplicial complex, $K$, as following: For any $i$, $T_i\subseteq K$; For any two vertices $v_1,v_2\in V$, if they are not neighbors, add an edge connecting them, whose length is their Euclidean distance in the Euclidean plane $\mathbb{E}$. See Figure \ref{fig:MCCPSpace} for the complex constructed from the input in Figure \ref{fig:MCCPPts} (Left). Please note that although in the figure, the embedding of $K$ in $\mathbb{R}^3$ has self-intersection, the simplicial complex $K$ can be studied without embedding, and thus, is well defined.

For the constructed complex $K$, we ask Opt-Diam to localize the 1-dimensional class $\sum_{i=1}^m h_i$, where $h_i$ is the only 1-dimensional class carried by the tube $T_i$. We need a cycle to represent it as the input for Opt-Diam. We use $z_0=\sum_{i=1}^m z_{i_0}$, where $z_{i_0}$ is the 1-cycle whose vertices are $v_{i_0}$, $v'_{i_0}$ and $v''_{i_0}$. $v_{i_0}$ is an arbitrary vertex in $S_i$. 

Next, we construct a cover $C$ from the solution of Opt-Diam, $z$, and show that $C$ is the solution of MCCP. We construct an intermediate vertex set $C_0\subseteq V$ as follows. A vertex $v$ belongs to $C_0$ if and only if any of $v_i$, $v'_i$ and $v''_i$ belongs to the vertex set of $z$, $\vertex(z)$. 
The solution $z$ is in the form $\sum_{i=1}^m z_i$, where $z_i$ represents class $h_i$. Therefore, $C_0$ has nonempty intersection with each vertex set $S_i$. 
We compute the cover $C$ by picking one vertex from each $S_i\cap C_0$.

Within the simplicial complex, $\diam(C)=\diam(C_0)$ and $|\diam(C_0)-\diam(z)|\leq 2\epsilon$. Furthermore, $C$ has the same diameter in the simplicial complex, $K$, and in the Euclidean plane, $\mathbb{E}$. Since $\epsilon$ is arbitrarily small, we can see that $C$ is the cover with the minimal diameter in the Euclidean plane, and thus, is the solution of MCCP.
%We produce an input for Opt-Diam: Let the simplicial complex be $K$. Pick an arbitrary cover $C_0=\{v_{i_0} \mid v_{i_0}\in S_i, i=1,..,m\}$. Let the 1-cycle $z_0=\sum_{i=1}^m z_{i_0}$ represent the 1-dimensional homology class we want to localize, where $z_{i_0}$ is the cycle whose vertices are $v_{i_0}$, $v_{i_0}'$ and $v_{i_0}''$.
%
%The only thing left is showing that a solution of Opt-Diam with the produced input leads to a solution of MCCP. Since the tube is very thin, we can see that the solution is in the form $\sum_{i=1}^m z_{i_1}$, where $z_{i_1}$ is the cycle whose vertices are $v_{i_1}$, $v_{i_1}'$ and $v_{i_1}''$. This solution corresponds to a cover $C_1=\{v_{i_1} \mid v_{i_1}\in S_i, i=1,..,m\}$, which is the solution of MCCP.
\qed

\subsection{Radius}
\label{sec:rad}
This section needs more notations about the discrete geodesic distance and the geodesic balls. We define the {\it discrete geodesic distance} from a vertex $p\in \vertex(K)$, $f_p:\vertex(K) \rightarrow \mathbb{Z}$, as follows. For any vertex
$q\in \vertex(K)$, $f_p(q)=\dist(p,q)$. For any simplex $\sigma \in K$, $f_p(\sigma)$ is the maximal function value of the vertices of $\sigma$, $f_p(\sigma)=\max_{q\in \vertex(\sigma)}f_p(q)$. 
Finally, we define a geodesic ball $B_p^r$, $p\in \vertex(K)$, $r\ge 0$, 
as the subset of $K$, $B_p^r = \{ \sigma \in K \mid f_p(\sigma) \le r \}$. It is straightforward to show that these subsets are in fact subcomplexes.

\begin{figure}[hbtp]
    \centerline{
    \begin{tabular}{c|c|c}
    \includegraphics[width=0.25\textwidth]{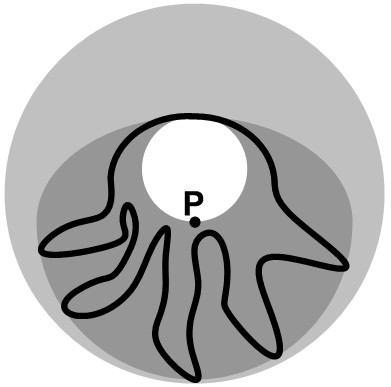} &
		\includegraphics[width=0.25\textwidth]{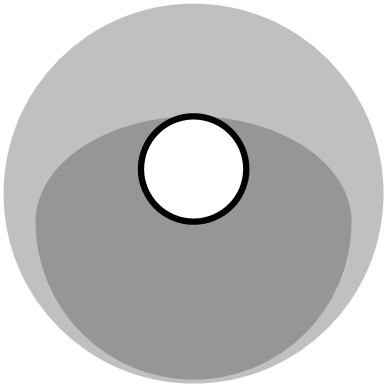} &
		\includegraphics[width=0.33\textwidth]{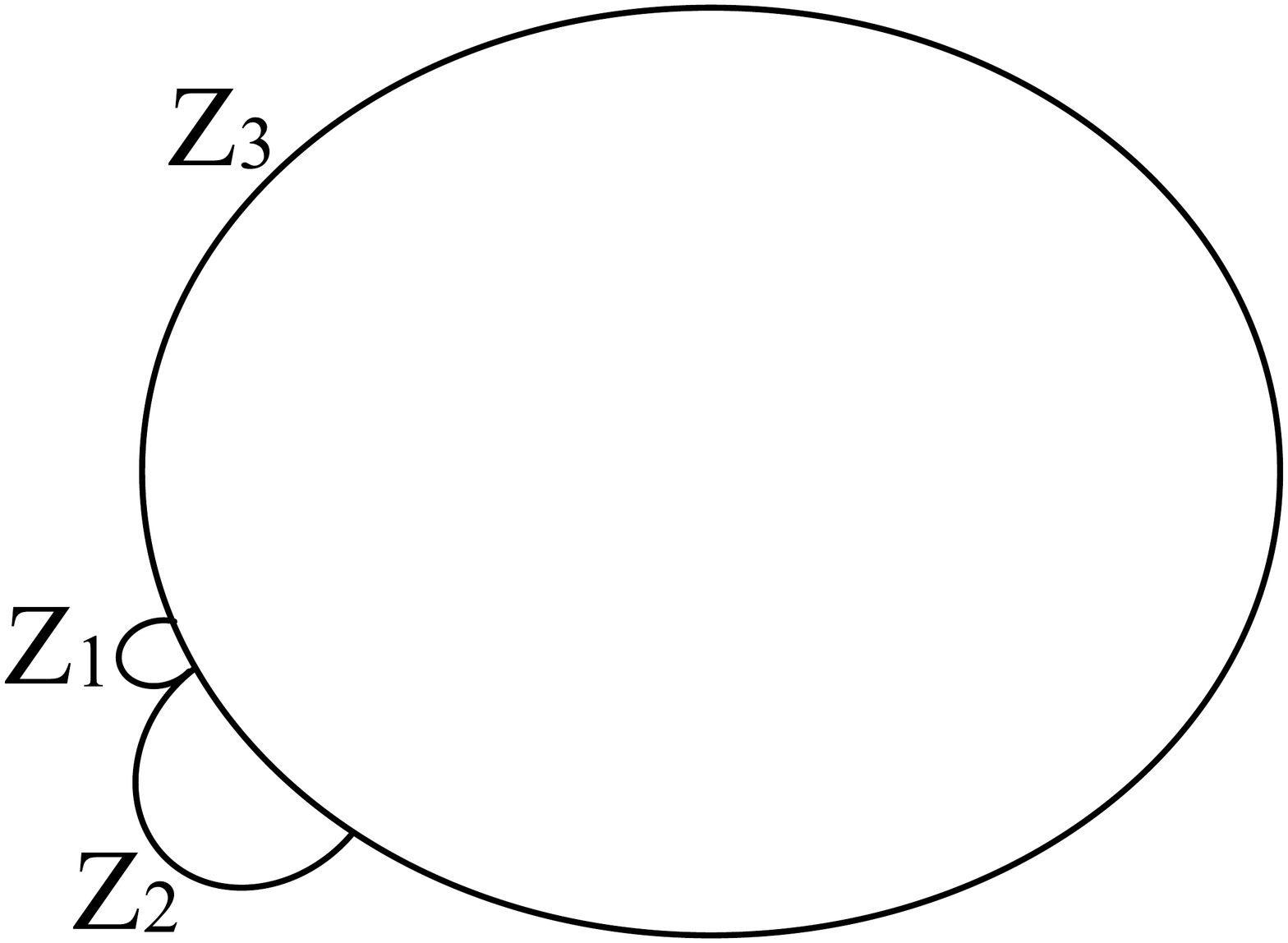}
    \end{tabular}}
    \vspace{-0in}
    \caption{\small\bf Left: the cycle with the minimal radius, $\zOptR$. 
    \newline Center: the cycle with the minimal diameter, $\zOptD$.
    \newline Right: Unstable Optimal Homology Basis. Used in Section \ref{sec:stabilityBasis}}
    \label{fig:wiggle}
\end{figure}

The third option of the objective function is the radius.
\begin{definition}[Radius]
The radius of a cycle is the radius of the smallest geodesic ball carrying it, formally,
\begin{equation*}
\rad(z)=\min_{p\in \vertex(K)}\max_{q\in \vertex(z)}\dist(p,q),
\end{equation*}
where $\vertex(K)$ and $\vertex(z)$ are the sets of vertices of the given simplicial complex $K$ and the cycle $z$, respectively.
\end{definition}
The representative cycle with the minimal radius, denoted as $\zOptR$, is the same as the localized-cycle defined in the companion paper \cite{ChenF1}. Intuitively, $\zOptR$ is the cycle whose vertices are as close to a vertex of $K$ as possible. However, $\zOptR$ may not necessarily be concise enough in intuition. It may wiggle a lot while still being carried by the smallest geodesic ball carrying the class. See Figure \ref{fig:wiggle} (Left), in which we localize the only nontrivial homology class of an annulus (the light gray area). The dark gray area is the smallest geodesic ball carrying the class, whose center is $p$. Note that the geodesic ball of the annulus may not seems like a disc in the embedded Euclidean plane. Besides, the cycle with the minimal diameter (Figure \ref{fig:wiggle} (Center)) avoids this wiggling problem and is concise in intuition. This in turn justifies the choice of diameter in Section \ref{sec:diam}.\footnote{This figure also illustrates that the radius and the diameter of a cycle are not strictly related. For the cycle $\zOptR$ in the left, its diameter is twice of its radius. For the cycle $\zOptD$ in the center, its diameter is equal to its radius.} 

In this section, we first show in Theorem \ref{thm:radApxDiam} that $\zOptR$ is a 2-approximation of $\zOptD$. Second, we show that $z_r$ can be computed in polynomial time.
\begin{theorem}
\label{thm:radApxDiam}
$\diam(z_r) \leq 2\diam(z_d)$.
\end{theorem}
\proof
First, the triangle inequality of the geodesic distance suggests that for any two vertices of $z_r$, $p_1$ and $p_2$, their geodesic distance is $\dist(p_1,p_2)\leq \dist(p_1,p_0)+\dist(p_0,p_2) \leq 2 \rad(z_r)$,
where $p_0$ is the center of the smallest geodesic ball carrying the cycle $z_r$ and the class. 
This implies that the diameter of $z_r$ is no greater than twice of its radius.

Second, the diameter of $z_d$ is no less than its radius. To see this, pick up a geodesic ball centered at any vertex of $z_d$ with radius $\diam(z_d)$. This ball carries $z_d$. 
Finally, $\diam(z_r)\leq 2\rad(z_r)\leq 2\rad(z_d)\leq 2\diam(z_d)$.
\qed
This bound is a tight bound. In Figure \ref{fig:wiggle}, the cycle in the left is $z_r$ whose diameter is twice of the radius of the dark gray geodesic ball. The cycle in the center is $z_d$ whose diameter is the same as the radius of the ball. We have $\diam(z_r)=2\diam(z_d)$.

\begin{theorem}
We can compute $\zOptR$ in polynomial time.
\end{theorem}
\proof
The proof is based on a polynomial algorithm. In the companion paper \cite{ChenF1}, we computed $z_r$ for each class in the optimal homology class. Based on this algorithm, we devise a polynomial algorithm computing $z_r$ for an arbitrary class. Because of the space limitation, we put the algorithm in Appdendix \ref{apx:radAlg}.
\qed
%%%%%%%%%%%%%%%%%%%%%%%%%%%%%%%%%%%%%%%%%%%%%%%%%%%%%%%%%%%%%%%%%%%%%%%%%%%%%%%%%%%%%%%%
\section{Stability Result.}
\label{sec:stability}
In this section, we prove that our measurement of homology in the companion paper \cite{ChenF1} is stable: small changes of the geometry of the space imply small changes of our measurement.
We define a change of the geometry of the space as a change of the metric in the space. We measure this change by measuring the $L_{\infty}$-norm difference of geodesic distance functions before and after the change. 
%We reuse the discrete geodesic distance function $f_p$ we used in Section \ref{sec:rad}, allowing the function value to be real values.\footnote{There may be issues about computing the discrete geodesic distance function. We circumvent these issues by assuming we are given the pairwise distance between all vertices. The distance function, $f_p$, on general dimensional simplices and the geodesic ball are defined in the same way as in Section \ref{sec:rad}.} 
To facilitate the proof, we assume that during the change, the simplicial complex remains the same except for the discrete geodesic distance. Formally, we quantify the change of the geometry as 
\begin{equation}
\epsilon=\max_{p\in \vertex(K)}|f_p^1-f_p^2|_{\infty}, 
\label{eqn:epsilon}
\end{equation}
where $f_p^1$ and $f_p^2$ are the discrete geodesic distance functions before and after the change.

Recall that in the companion paper, our measurement of homology involves two concepts, namely, the size of a class and the optimal homology basis. we defined the size of a homology class $h$ as the radius of the smallest geodesic ball carrying $h$, formally,
\begin{eqnarray*}
S(h)=\min_{p\in \vertex(K)} r_{f_p}(h),
\end{eqnarray*}
where $r_{f_p}(h)$ is the smallest value $t$ such that the geodesic ball $f_p^{-1}(-\infty,t]$ carries $h$. A geodesic ball carried $h$ if and only if it carries one of the cycles of $h$. Based on this size definition, we defined the optimal homology basis $\mathcal{H}$ as the set of $\beta$ linearly independent homology classes whose size have the minimal sum, where $\beta$ is the Betti number. Matroid theory suggests that $\mathcal{H}$ can be computed using a greedy algorithm: sort all the nontrivial classes in a monotonically increasing order according to their size. Check the smallest classes one by one. Pick each class that is not linearly dependent on what we have chosen, until $\beta$ classes are collected.

In this section, we prove the stability of our measurement by showing that (1) for a single homology class, the size is stable; and (2) for the whole homology group, although the optimal homology basis is not stable, the group structure filtered by the size is stable.

\subsection{A Single Class}
For a single homology class, the size measure remains stable. Denote $S^1(h)$ and $S^2(h)$ as the size of class $h$ before and after the change (computed using $f_p^1$ and $f_p^2$, respectively). We have the following theorem.
\begin{theorem}
\label{thm:classStable}
$|S^1(h)-S^2(h)|\leq \epsilon$, where $\epsilon$ is the upperbound of the geometry change as defined in Equation (\ref{eqn:epsilon}).
\end{theorem}
\proof
We show that for any specific vertex $p$, $|r_{f_p^1}(h)-r_{f_p^2}(h)|\leq \epsilon$. This leads to the fact that $S^1(h)=\min_{p\in \vertex(K)} r_{f_p^1}(h)$ and $S^2(h)=\min_{p\in \vertex(K)} r_{f_p^2}(h)$ differ in no more than $\epsilon$.

For any vertex $p$, denote $r_1$ and $r_2$ as $r_{f_p^1}(h)$ and $r_{f_p^2}(h)$ for short. According to the definition, they are the radii of the smallest geodesic balls carrying $h$ computed using the geodesic distance $f_p^1$ and $f_p^2$, respectively. For any simplex $\sigma$ in the ball $B_{p}^{r_1}$ calculated using $f_p^1$, $f_{p}^1(\sigma)\leq r_1$, and thus $f_{p}^2(\sigma)\leq f_{p}^1(\sigma)+\epsilon \leq r_1+\epsilon$. This means that the ball $B_{p}^{r_1}$ calculated using $f_p^1$ is a subcomplex of the ball $B_{p}^{r_1+\epsilon}$ calculated using $f_p^2$. Therefore, according to the definition of $r_2$, it is no greater than $r_1+\epsilon$. Similarly, we can see that $r_1$ is no greater than $r_2+\epsilon$.
\qed

\subsection{The Homology Group}
\label{sec:stabilityBasis}
Since the size of different classes can be very close, the optimal homology basis is not stable. For example, in Figure \ref{fig:wiggle} (Right), either $\{[z_1],[z_2],[z_3]\}$ or $\{[z_1],[z_2],[z_3]+[z_1]\}$ can be the optimal homology basis for little geometry changing, because the sizes of $[z_3]$ and $[z_1]+[z_3]$ are quite close. However, there is still some stability property in the homology group structure if we filter it with the class size. More specifically, the subgroup generated by small homology classes remains stable. For example, in Figure \ref{fig:wiggle} (Right), although the optimal homology basis is unstable, the subgroup generated by the two smaller classes in the optimal homology basis will always be the one generated by $[z_1]$ and $[z_2]$.

We formalize this stability by defining the subgroup filtration of a topological space and the distance between two such filtrations. A subgroup filtration is a sequence of subgroups of the homology group generated by subsets of the optimal homology basis filtered by the class size. A formal definition is as follows.
\begin{definition}[Subgroup Filtration]
Given an optimal homology basis $\mathcal{H}=\{h_1,h_2,...,h_{\beta}\}$, where we assume $S(h_i)\leq S(h_{i+1})$, a \emph{subgroup filtration} is a sequence of subgroups of the homology group, $\mathcal{X}=\{\psi_0,\psi_1,\psi_2,...,\psi_{\beta}\}$, where $\psi_i=\spanSpace(h_1,h_2,...,h_i)$ is the subgroup generated
\footnote{Since here the coefficient ring $R=\mathbb{Z}_2$, the homology group and all its subgroups are vector spaces. Therefore, we use the notation $\psi_i=\spanSpace(h_1,h_2,...,h_i)$ when we say $h_1,h_2,...,h_i$ generates $\psi_i$.}
by the classes $h_1$, $h_2$, ..., $h_i$.
\label{def:subgroupFiltration}
\end{definition}
Obviously, the subgroup filtration is a sequence of subgroups of $\mathsf{H}(K)$ with a nested structure $\psi_i\subseteq \psi_{i+1}$, $\psi_0=\emptyset$ and $\psi_{\beta}=\mathsf{H}(K)$. For convenience, we denote the size of a subgroup, $\psi_i$, as the size of the largest class in the optimal homology basis generating $\psi_i$, formally, $S(\psi_i)=S(h_i)$. Please note that $S(\psi_i)$ is not the size of the largest class in $\psi_i$. Next, we define the distance function between two subgroup filtrations $\mathcal{X}^1$ and $\mathcal{X}^2$, which requires the definition of the projection of one subgroup in one filtration onto the other filtration.
\begin{definition}[Projection]
Given two subgroup filtrations of a same homology group $\mathcal{X}^1=\{\psi_0^1,\psi_1^1,\psi_2^1,...,\psi_{\beta}^1\}$ and $\mathcal{X}^2=\{\psi_0^2,\psi_1^2,\psi_2^2,...,\psi_{\beta}^2\}$, define the \emph{projection} of $\psi_i^1$ onto $\mathcal{X}^2$ as the first subgroup in $\mathcal{X}^2$ that carries $\psi_i^1$, formally,
\begin{eqnarray*}
\proj(\psi_i^1,\mathcal{X}^2)=\psi_j^2, \quad {\rm s.t.} j=\min_{\psi_i^1\subseteq \psi_k^2}k.
\end{eqnarray*}
\end{definition}
\begin{definition}[Distance]
Define the \emph{distance} between $\mathcal{X}^1$ and $\mathcal{X}^2$ as the maximal difference between the sizes of any subgroup in $\mathcal{X}^1$ or $\mathcal{X}^2$ and its projection onto the other filtration, formally,
\begin{eqnarray*}
\dist(\mathcal{X}^1,\mathcal{X}^2)=\max\{\max_i|S^1(\psi_i^1)-S^2(\proj(\psi_i^1,\mathcal{X}^2))|, \max_i|S^2(\psi_i^2)-S^1(\proj(\psi_i^2,\mathcal{X}^1))|\}.
\end{eqnarray*}
\end{definition}
%Please note a given homology class (or a subgroup) has two different sizes before and after the change. We denote $S^1(h)$ and $S^2(h)$ as the two sizes (or $S^1(\psi)$ and $S^2(\psi)$).

Let $\mathcal{X}^1$ and $\mathcal{X}^2$ be the subgroup filtrations of the original space and the one after the change. We can prove the following stability result.
\begin{theorem}
\label{thm:groupStable}
\begin{eqnarray*}
\dist(\mathcal{X}^1,\mathcal{X}^2)\leq \epsilon = \max_{p\in \vertex(K)} |f_p^1-f_p^2|_{\infty}.
\end{eqnarray*}
\end{theorem}
To facilitate the proof, we state the following two facts about any optimal homology basis $\mathcal{H}=\{h_1,h_2,\cdots,h_{\beta}\}$ (sorted according to their size), which follows the greedy algorithm of computing $\mathcal{H}$.
\begin{itemize}
\item {\bf Fact 1:} Any class in the basis, $h_j\in \mathcal{H}$, cannot be linearly dependent on classes whose size are smaller than $S(h_j)$;
\item {\bf Fact 2:} For any homology class $h$ and real number $s_0$, $S(h)\leq s_0$ implies that there exists a $k\leq \beta$, such that $h\in \spanSpace(h_1,h_2,\cdots,h_k)$ and $S(h_k)\leq s_0$.
\end{itemize}
With these facts, we prove Theorem \ref{thm:groupStable} as follows.
\proof
It suffices to show that for any subgroup $\psi_i^1$, $|S^1(\psi_i^1)-S^2(\psi_j^2)|\leq \epsilon$, where $\psi_j^2$ is the projection, $\proj(\psi_i^1,\mathcal{X}^2)$. Denote the optimal homology bases before and after the change as $\mathcal{H}^1=\{h_1^1,h_2^1,...,h_{\beta}^1\}$ and $\mathcal{H}^2=\{h_1^2,h_2^2,...,h_{\beta}^2\}$, both of which are sorted by the class size. We know that $\psi_i^1=\spanSpace(h_1^1,h_2^1,...,h_i^1)$ and $\psi_j^2=\spanSpace(h_1^2,h_2^2,...,h_j^2)$. We will use the following two facts which follows the construction of the optimal homology basis.

We first prove by contradiction that 
\begin{eqnarray}
S^2(\psi_j^2)\geq S^1(\psi_i^1)-\epsilon.
\label{eqn:groupStableE1}
\end{eqnarray} 
Suppose $S^2(\psi_j^2)<S^1(\psi_i^1)-\epsilon$. Any class in the set $\{h_1^2,h_2^2,\cdots,h_j^2\}$ has its $S^2$ size smaller than $S^1(h_i^1)-\epsilon$, 
%formally, for any $k\leq j$, 
%\begin{eqnarray*}
%S^2(h_k^2)\leq S^2(h_j^2)<S^1(\psi_i^1)-\epsilon=S^1(h_i^1)-\epsilon.
%\end{eqnarray*}
and thus, according to Theorem \ref{thm:classStable} has its $S^1$ size smaller than
$S^1(h_i^1)$.
Based on the definition of the projection, $\psi_i^1\subseteq \psi_j^1$. The class $h_i^1$ is linearly dependent on classes $h_1^2,h_2^2,\cdots,h_j^2$, all of which have their $S^1$ size smaller than $S^1(h_i^1)$. This contradicts Fact 1. Equation (\ref{eqn:groupStableE1}) is proved.

Second, we show that
\begin{eqnarray}
S^2(\psi_j^2)\leq S^1(\psi_i^1)+\epsilon.
\label{eqn:groupStableE2}
\end{eqnarray} 
For any $l\leq i$, $S^1(h_l^1)\leq S^1(h_i^1)$. According to Theorem \ref{thm:classStable}, 
\begin{eqnarray*}
S^2(h_l^1)\leq S^1(h_l^1)+\epsilon \leq S^1(h_i^1)+\epsilon = S^1(\psi_i^1)+\epsilon.
\end{eqnarray*}
Based on Fact 2, for all such $l$, there exists $k$, such that $h_l^1\in \spanSpace(h_1^2,h_2^2,\cdots,h_k^2)$ and $S^2(\psi_k^2)=S^2(h_k^2)\leq S^1(\psi_i^1)+\epsilon$. Therefore, $\psi_i^1\subseteq \psi_k^2$. Based on the definition of the projection, $\psi_j^2$ has its size no greater than $\psi_k^2$. Equation (\ref{eqn:groupStableE2}) is proved.
\qed
\section{Future Problems}
%In this paper, we address the problem of localizing homology classes with a cycle with the minimal size. We explored three different choices of the size of cycles, namely, the volume, the diameter and the radius. We Prove that computing the cycle with the minimal volume or diameter is NP-hard. We also show that although the cycle with the minimal radius may not be concise in intuition, it is a 2-approximation of the cycle with the minimal volume. A polynomial algorithm is provided to compute the cycle with the minimal radius.
We found the following related problems to be interesting and deserving further study.

{\bf (1)} We prove that localizing any given class with the cycle with the minimal volume is NP-hard. However, notice that Erickson and Whittlesey \cite{EricksonW05} showed in low dimension that we can localize the shortest homology basis in polynomial time. An interesting question is then can we in general localize the basis with the minimal volume sum in polynomial time? What about the diameter?

{\bf (2)} We restricted the stability proof to the case when the geometry change is a homeomorphism (the simplicial complex remains unchanged). Is it possible to extend this to incorporate topological changes? More specifically, is the measurement still stable even if small homology classes are destroyed and created during the change?

\end{spacing}
\bibliographystyle{abbrv}
\bibliography{topologyBib}

\begin{spacing}{0.95}
\newpage\appendix
\section{A Polynomial Algorithm to Compute $\zOptR$, the Cycle with the Minimal Radius}
\label{apx:radAlg}
In the companion paper, a polynomial algorithm is provided to compute $\zOptR$ for classes in the optimal homology basis. We generalize this algorithm to compute $\zOptR$ for any given homology class. Please refer to \cite{ChenF1} for the correctness of this algorithm. 

First, we need the smallest geodesic ball carrying the class we want to localize. The procedure {\sf Bmin($K$,$z_0$)} in Algorithm \ref{alg:Bmin} computes this ball, which uses the procedure {\sf Contain-Cycle($K$,$K_0$,$z_0$)} (See Algorithm \ref{alg:containCycle}) to detect whether a subcomplex $K_0$ carries any cycle of the class to localize.
\begin{algorithm}[h!]
 \caption{\small\bf {\sf Bmin($K$,$z_0$)}}
 \label{alg:Bmin}

 \begin{algorithmic}[1]
   \GOAL   Find the smallest geodesic ball, $B_{min}$, carrying any cycle of $h$.

    \INPUT $K$: the given simplicial complex.\\
    			 $z_0$: a cycle representing $h$.

    \OUTPUT $p_{min}$ and $r_{min}$: the center and radius of the computed ball $B_{min}$.
		
		\FOR {$p\in \vertex(K)$}
		\STATE 	r=0
		\REPEAT
		\STATE		r=r+1
		\UNTIL {{\sf Contain-Cycle($K$,$B_p^r$,$z_0$)}}
		\IF {$r < r_{min}$}
				\STATE $r_{min}=r$
				\STATE $p_{min}=p$
		\ENDIF
		\ENDFOR
 \end{algorithmic}
\end{algorithm}

\begin{algorithm}[h!]
 \caption{\small\bf {\sf Contain-Cycle($K$,$K_0$,$z_0$)}}
 \label{alg:containCycle}

 \begin{algorithmic}[1]
   \GOAL   test whether $K_0$ carries any representative cycle of the class $h=[z_0]$.

    \INPUT $K$: the given simplicial complex.\\
    			 $K_0$: the subcomplex.\\
    			 $z_0$: a cycle representing $h$.

    \OUTPUT Boolean.
		
		\STATE $\hat{Z}_d=[\partial_{d+1},z_0]$
		\STATE compute $\partial_{d+1}^{K\backslash K_0}$ and $\hat{Z}_d^{K\backslash K_0}$ by picking up
		rows of $\partial_{d+1}$ and $\hat{Z}_d$ whose corresponding simplices do not belong to $K_0$
    \IF{ $\rank(\hat{Z}_d^{K\backslash K_0})-\rank(\partial_{d+1}^{K\backslash K_0})\neq 1$ }
			\STATE return true
		\ELSE
			\STATE return false
		\ENDIF
 \end{algorithmic}
\end{algorithm}

Second, when $B_{min}$ is computed, we compute a cycle of $h$ carried by $B_{min}$ as follows.
Solve the equation system $\partial_{d+1}^{K\backslash B_{min}}\gamma+z_0^{K\backslash B_{min}}=0$, where $\partial_{d+1}^{K\backslash B_{min}}$ is the matrix formed by the rows of $\partial_{d+1}$ whose corresponding simplices do not belong to $B_{min}$, $z_0^{K\backslash B_{min}}$ is the vector formed by the entries of $z_0$ whose corresponding simplices do not belong to $B_{min}$. Using the solution $\gamma$, $z_0+\partial_{d+1}\gamma$ is a cycle of $h$ carreid by $B_{min}$, and thus is the desired $\zOptR$.
\end{spacing}
\end{document}